\documentclass[prb,twocolumn,superscriptaddress,showpacs]{revtex4}
\usepackage{amsmath,amssymb,graphicx,color,bm,epstopdf}

\begin{document}
\title{Persistent current induced by vacuum fluctuations in a quantum ring}

\author{O. V. Kibis}
\affiliation{Department of Applied and Theoretical Physics,
Novosibirsk State Technical University, Karl Marx Avenue 20,
Novosibirsk 630073, Russia}

\author{O. Kyriienko}
\affiliation{Science Institute, University of Iceland, Dunhagi-3,
IS-107, Reykjavik, Iceland} \affiliation{Division of Physics and
Applied Physics, Nanyang Technological University 637371,
Singapore}

\author{I. A. Shelykh}
\affiliation{Science Institute, University of Iceland, Dunhagi-3,
IS-107, Reykjavik, Iceland} \affiliation{Division of Physics and
Applied Physics, Nanyang Technological University 637371,
Singapore}


\begin{abstract}
We study theoretically interaction between electrons in a quantum
ring embedded in a microcavity and vacuum fluctuations of
electromagnetic field in the cavity. It is shown that the vacuum
fluctuations can split electron states of the ring with opposite
angular momenta. As a consequence, the ground state of electron
system in the quantum ring can be associated to nonzero electric
current. Since a ground-state current flows without dissipation,
such a quantum ring gets a magnetic moment and can be treated as
an artificial spin.
\end{abstract}

\maketitle


\section{Introduction} The interaction between light and matter
represents an important part of the modern physics, from both
fundamental and applied point of view. In particular, the vast
fundamental research is devoted to studies of electromagnetic
vacuum.\cite{Milonni} Being one of the cornerstones of quantum
electrodynamics, observing alteration of atom levels due to vacuum
fluctuations (the Lamb shift) \cite{Lamb_47,Bethe_47,Scully_10}
and attraction between conducting plates caused by radiational
pressure of virtual photons (the Casimir effect)
\cite{Casimir,Fialkovsky_11,Sernelius_11} have led to deeper
understanding of the electromagnetic field. However, the influence
of vacuum fluctuations is usually minor in the non-relativistic
physics and is only accessible in state-of-the-art experiments.
Thus, the question of proposal for macroscopically observable
effects caused by electromagnetic fluctuations of vacuum is still
open.\cite{Jaffe}

The physics of light-matter interaction contains the wide range of
topics, namely cavity quantum
electrodynamics,\cite{Dutra,Review_CQED} laser
physics,\cite{LaserRev1,LaserRev2}
polaritonics,\cite{KavokinBook,PolaritonDevices} etc. While most
of the topics assume the emission and absorption of real photons
by particles in a solid, the light-matter interaction is not only
restricted to this case. For instance, the electronic states can
be ``dressed'' by photons, changing the energy spectrum of
electron-photon system, while photon absorption is
prohibited.\cite{Cohen-Tannoudji_b98} This is the essence of
dynamic Stark effect \cite{Autler} studied before for various
systems (see, e.g.,
Refs.~[\onlinecite{ScullyZubairy,Koch,Faist,Wu,Kibis_10,Vidar,Snoke}]).
However, previously proposed experimental configurations require
the source of \emph{real} photons which are directly detectable
quanta of electromagnetic field. In this paper we study the
dynamic Stark effect induced by \emph{virtual} photons --- vacuum
fluctuations of electromagnetic field confined in a resonator  ---
for the particular case of electron states in a quantum ring
embedded in the optically chiral resonator. Due to the
vacuum-induced splitting of electron energy levels with opposite
angular momenta, the ground state of electron system in the ring
can be associated to nonzero angular momentum. As a consequence, a
ground-state dissipationless electric current (persistent current)
appears. It should be stressed that the discussed phenomenon
differs conceptually from persistent currents in Aharonov-Bohm
quantum rings,\cite{Buttiker,Mailly} where the ground-state
dissipationless current is caused by an external magnetic flux
through the ring. Thus, we present the theory of significant novel
mechanism of dissipationless electron transport, where physics of
nanostructures and quantum electrodynamics meet.

\section{The model} 
\begin{figure}
\includegraphics[width=0.7\linewidth]{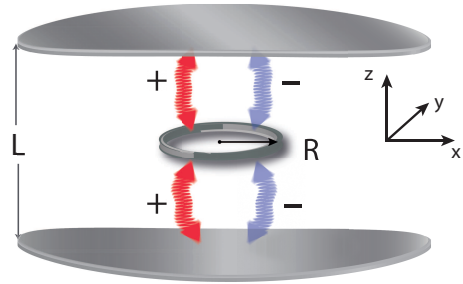}
\caption{(Color online) Sketch of the system. A quantum ring of
radius $R$ is placed inside a planar cavity of length $L$. The
arrows with signs $+$ and $-$ correspond to clockwise and
counterclockwise circularly polarized virtual photons.}
\label{Fig1}
\end{figure}

We consider the problem of interaction between
an electron in a one-dimensional quantum ring and a virtual photon
mode of a planar resonator (microcavity). The geometry of the
system is shown in Fig.~\ref{Fig1} and represents the conducting
ring of radius $R$ placed inside the resonator with the cavity
length $L$.
The Hamiltonian of the considered electron-photon system has the
form
\begin{equation}
\hat{\cal{H}}=\hat{\cal{H}}_{el}+\hat{\cal{H}}_{ph}+\hat{\cal{H}}_{int},
\label{Htot}
\end{equation}
where $\hat{\cal{H}}_{el}$ is the Hamiltonian of an electron in
the ring, $\hat{\cal{H}}_{ph}$ is the Hamiltonian of a photonic
mode in the cavity, and $\hat{\cal{H}}_{int}$ is the Hamiltonian
of electron-photon interaction.

The electron Hamiltonian is given by the expression
\begin{equation}
\hat{\cal{H}}_{el}= \frac{\hbar^2 \hat{l}^{2}_{z}}{2 m_{e}R^2},
\label{Hel}
\end{equation}
where $m_{e}$ is the effective mass of electron in the ring, $R$
is the radius of the ring, $\hat{l}_{z}=-i
\partial/\partial \varphi$ is operator of dimensionless electron angular momentum, and $\varphi$ is
the angular coordinate of electron in the ring.

The photon Hamiltonian, accounting for both clockwise ($\lambda =
+ $) and counterclockwise ($\lambda = -$) circular polarizations,
reads as
\begin{align}
\label{Hph}
\hat{\cal{H}}_{ph}&=\sum_{\mathbf{q},\eta,\lambda} \hbar\omega_{\mathbf{q},\eta,\lambda} \hat{a}^\dagger_{\mathbf{q},\eta,\lambda} \hat{a}_{\mathbf{q},\eta,\lambda} \\
&+\sum_{\mathbf{q},\eta} \frac{\hbar\Omega_{LT}(\mathbf{q})}{2}
(\hat{a}^\dagger_{\mathbf{q},\eta,+} \hat{a}_{\mathbf{q},\eta,-} +
\hat{a}^\dagger_{\mathbf{q},\eta,-} \hat{a}_{\mathbf{q},\eta,+}),
\notag
\end{align}
where $\hat{a}^\dagger_{\mathbf{q},\eta,\lambda}$ and
$\hat{a}_{\mathbf{q},\eta,\lambda}$ are creation and annihilation
operators for cavity photons with polarizations $\lambda=\pm$ and
wave vectors $(\mathbf{q},q_z)$. Here $\mathbf{q}$ is the in-plane
component of photon wave vector in the cavity, $q_z=\eta\pi/L$ is
the quantized $z$ component of photon wave vector in the cavity,
and $\eta=1,2,3,...$ is the number of photon mode in the cavity.
Correspondingly, the first term in Eq. (\ref{Hph}) describes the
energy of cavity modes with dispersions given by
\begin{equation}\label{omega}
\omega_{\mathbf{q},\eta,\pm} = c_\pm\sqrt{q^2 + q_{z}^2},
\end{equation}
where $c_\pm=c/n_\pm$ are the speeds of light with clockwise and
counterclockwise circular polarizations, and $n_\pm$ are the
refractive indices for clockwise ($\lambda=+$) and
counterclockwise ($\lambda=-$) polarized light. In what follows we
will consider the case of chiral resonator. Thus, in general,
$c_+\neq c_-$. As to the second term in Eq. (\ref{Hph}), it
describes the energy splitting between photon modes with different
polarizations in a microcavity (longitudinal-transverse
splitting).\cite{ShelykhRev} The exact form of the
longitudinal-transverse splitting function
$\Omega_{LT}(\mathbf{q})$ depends on the construction of the
resonator, but in majority of cases it can be approximated by the
simple formula $\Omega_{LT}(\mathbf{q})=\hbar q^2/2\mu$, where
$\mu=m_{TE}m_{TM}/(m_{TM}-m_{TE})$, $m_{TE}$ and $m_{TM}$ are the
effective masses of cavity photons with TE and TM polarizations,
respectively. For a typical microcavity structure, they can be
found as $m_{TE}=3.68\times 10^{-4} m_0$ and $m_{TM}=3.62\times
10^{-4} m_0$, where $m_0$ is the mass of free
electron.\cite{Kalita} The presence of the longitudinal-transverse
splitting affects the polarization of eigenmodes of the planar
cavity, as it will be discussed below.

Taking into account one-dimensional geometry of the quantum ring,
the interaction Hamiltonian has the form \cite{Kibis11}
\begin{equation}
\hat{\cal{H}}_{int}=-eR\sum_{\mathbf{q},\eta,\lambda} \int
\hat{\mathbf{E}}_{\mathbf{q},\eta,\lambda}
\mathbf{t}(\varphi)d\varphi, \label{Hint}
\end{equation}
where the indefinite integral should be treated as an
antiderivative of the subintegral function. Here
$\mathbf{t}(\varphi)=-\mathbf{e}_{x}\sin \varphi +
\mathbf{e}_y\cos \varphi$ is the unit tangent vector to the ring,
$\mathbf{e}_{x}$ and $\mathbf{e}_{y}$ are the in-plane Cartesian
unit vectors, the operator of the electric field in the cavity is
\begin{equation}
\hat{\mathbf{E}}_{\mathbf{q},\eta,\lambda}=i\sqrt{\frac{\hbar
\omega_{\mathbf{q},\eta,\lambda}}{2 \epsilon_{0}}}\left(
\hat{a}_{\mathbf{q},\eta,\lambda}
\mathbf{u}_{\mathbf{q},\eta,\lambda} -
\hat{a}^{\dagger}_{\mathbf{q},\eta,\lambda}
\mathbf{u}^{*}_{\mathbf{q},\eta,\lambda} \right), \label{E}
\end{equation}
eigenvectors of the cavity are given by the expression
\cite{ScullyZubairy}
\begin{equation}
\mathbf{u}_{\mathbf{q},\eta,\lambda}=\mathbf{e}_{\lambda,\mathbf{q}}\sqrt{\frac{2}{L
S}}\sin \left( \frac{\pi \eta z}{L} \right) e^{i\mathbf{q}\cdot
\mathbf{r}}, \label{u}
\end{equation}
$L$ is the cavity length, $S$ is the cavity area, $\mathbf{r}$ is
the in-plane radius vector, and $\mathbf{e}_{\lambda,\mathbf{q}}$
are the unit vectors of photon polarizations.

In order to describe the noninteracting electron-photon system in
the cavity, let us use the jointed electron-photon
space,\cite{KibisPRB} $|m,N_{\mathbf{q},\eta,\lambda}\rangle =
|m\rangle \otimes |N_{\mathbf{q},\eta,\lambda}\rangle$, which
indicates that the electromagnetic field is in a quantum state
with the photon occupation number
$N_{\mathbf{q},\eta,\lambda}=0,1,2,3...$ , and the electron is in
a quantum state with the wave function
$\psi_{m}(\varphi)=1/\sqrt{2 \pi}\exp(im\varphi)$, where $m=0,\pm
1,\pm 2...$ is the electron angular momentum along the ring axis.
It should be noted that polarizations of eigenmodes of the photon
Hamiltonian (\ref{Hph}) are, in general, elliptical and strongly
depend on in-plane photon wave vector $\mathbf{q}$, transforming
into circular polarization for $q\rightarrow 0$ and into linear
one for $q\rightarrow\infty$.\cite{ShelykhRev} These elliptically
polarized eigenmodes of the photon Hamiltonian (\ref{Hph}) can be
found using the Hopfield transformations:\cite{Haug}
\begin{align}
\label{Hopfield1}
\hat{a}_{\mathbf{q},\eta, 1}=\alpha_{\mathbf{q}}\hat{a}_{\mathbf{q},\eta,+}+ \beta_{\mathbf{q}}\hat{a}_{\mathbf{q},\eta,-},\\
\hat{a}_{\mathbf{q},\eta,
2}=\beta_{\mathbf{q}}\hat{a}_{\mathbf{q},\eta,+}-\alpha_{\mathbf{q}}\hat{a}_{\mathbf{q},\eta,-},
\label{Hopfield2}
\end{align}
where the Hopfield coefficients can be written as
\begin{align}
\label{alpha}
\alpha_{\mathbf{q}}=\frac{-\Omega_{LT}(\mathbf{q})}{\sqrt{\Omega_{LT}^2(\mathbf{q})+\left(\Delta_{\pm,\eta}(\mathbf{q})- \sqrt{\Delta_{\pm,\eta}^{2}(\mathbf{q}) +\Omega_{LT}^2(\mathbf{q})}\,\right)^2}},\\
\label{beta}
\beta_{\mathbf{q}}=\frac{\Delta_{\pm,\eta}(\mathbf{q})-
\sqrt{\Delta_{\pm,\eta}^{2}(\mathbf{q})+\Omega_{LT}^2(\mathbf{q})}}{\sqrt{\Omega_{LT}^2(\mathbf{q})+\left(\Delta_{\pm,\eta}(\mathbf{q})-
\sqrt{\Delta_{\pm,\eta}^{2}(\mathbf{q})
+\Omega_{LT}^2(\mathbf{q})}\,\right)^2}},
\end{align}
and $\Delta_{\pm,\eta}(\mathbf{q})= \omega_{\mathbf{q},\eta,+} -
\omega_{\mathbf{q},\eta,-}$. Correspondingly, eigenfrequencies of
the cavity photon modes are
\begin{align}
\label{omega1}
\omega_{\mathbf{q},\eta,1}= \frac{\omega_{\mathbf{q},\eta,+}+ \omega_{\mathbf{q},\eta,-}}{2} + \frac{1}{2} \sqrt{\Delta_{\pm,\eta}^2(\mathbf{q}) + \Omega_{LT}^2(\mathbf{q})},\\
\label{omega2} \omega_{\mathbf{q},\eta,2}=
\frac{\omega_{\mathbf{q},\eta,+}+ \omega_{\mathbf{q},\eta,-}}{2} -
\frac{1}{2} \sqrt{\Delta_{\pm,\eta}^2(\mathbf{q}) +
\Omega_{LT}^2(\mathbf{q})},
\end{align}
and the diagonalized photon Hamiltonian (\ref{Hph}) reads as
\begin{equation}
\hat{\cal{H}}_{ph}=\sum_{\mathbf{q},\eta,\lambda^{\prime}}
\hbar\omega_{\mathbf{q},\eta,\lambda^{\prime}}
\hat{a}^\dagger_{\mathbf{q},\eta,\lambda^{\prime}}
\hat{a}_{\mathbf{q},\eta,\lambda^{\prime}}, \label{H0}
\end{equation}
where $\lambda^{\prime}=1,2$ is the polarization index of the
above-mentioned elliptical basis. As a result, the energy spectrum
of the noninteracting electron-photon system in the cavity is
\begin{equation}
\varepsilon_{m,N_{\mathbf{q},\eta,\lambda^{\prime}
}}^{(0)}=\frac{\hbar^2 m^2}{2m_{e}R} +
N_{\mathbf{q},\eta,\lambda^{\prime}}\hbar
\omega_{\mathbf{q},\eta,\lambda^{\prime}}. \label{eps0}
\end{equation}

For the case of electromagnetic vacuum in the cavity, photon
occupation numbers in Eq.~(\ref{eps0}) are
$N_{\mathbf{q},\eta,\lambda^{\prime}}=0$. Considering the electron
interaction with the photon vacuum as a weak perturbation
described by the Hamiltonian (\ref{Hint}), we can apply the
conventional perturbation theory. Then the energy spectrum of
electron in the ring dressed by vacuum fluctuations is given by
\begin{align}
\notag \varepsilon_{m,0} =& \varepsilon_{m,0}^{(0)} +
\sum_{\mathbf{q},m^{\prime},\eta} \left(\frac{| \langle
m^{\prime},1_{\mathbf{q},\eta,1}|\hat{\cal{H}}_{int}|m,0\rangle|^{
2}}{\varepsilon_{m,0}^{(0)}-
\varepsilon_{m^{\prime},1_{\mathbf{q},\eta,1}}}\right. \\ &+
\left.\frac{| \langle
m^{\prime},1_{\mathbf{q},\eta,2}|\hat{\cal{H}}_{int}|m,0\rangle|^{2}}{\varepsilon_{m,0}^{(0)}-
\varepsilon_{m^{\prime},1_{\mathbf{q},\eta,2}}}\right).
\label{eps}
\end{align}
Writing the interaction Hamiltonian (\ref{Hint}) for the
elliptical polarizations $\lambda=1,2$ and assuming the ring to be
placed in the center of the cavity, the expression for the
electron energy spectrum (\ref{eps}) takes the final form (see
detailed derivation in Appendix A):
\begin{align}
\label{eps_final} &\varepsilon_{m,0} = \varepsilon_{m,0}^{(0)} +
\sum_{m^{\prime},\eta} \frac{e^2 R^2}{2\pi} \frac{1}{2
\epsilon_{0}L} \frac{1}{(m-m^{\prime})^2} \\ \notag & \times\Bigg(
\int\limits_{0}^{\infty} dq \frac{q\hbar \omega_{q,\eta,1}
(J_{m-m^{\prime}-1}^{2}(qR)\alpha_{q}^2 +
J_{m-m^{\prime}+1}^{2}(qR)
\beta_{q}^2)}{[\varepsilon_{R}(m^2-m^{\prime 2})-\hbar
\omega_{q,\eta,1}]} \\ \notag & + \int\limits_{0}^{\infty} dq
\frac{q\hbar \omega_{q,\eta,1}
(J_{m-m^{\prime}-1}^{2}(qR)\beta_{q}^2 +
J_{m-m^{\prime}+1}^{2}(qR)
\alpha_{q}^2)}{[\varepsilon_{R}(m^2-m^{\prime 2})-\hbar
\omega_{q,\eta,2}]} \Bigg),
\end{align}
where $\varepsilon_{R} = \hbar^2 /2m_e R^2$ is the characteristic
electron energy in the ring, and $\eta=1,3,5,...$ is odd integer.

\section{Discussion} It should be noted that the integrals in
Eq.~(\ref{eps_final}) are divergent. This divergency arises from
the accounting of infinite number of vacuum modes and has the same
origin as a formally infinite energy of vacuum state in the
cavity. However, the physically measurable quantity is not the
shift of electron energy levels but the splitting of them by
vacuum fluctuations. Particularly, the splitting of electron
energy levels with mutually opposite angular momenta $m$ and $-m$,
\begin{equation}\label{D}
\Delta \varepsilon = |\varepsilon_{m,0} - \varepsilon_{-m,0}|,
\end{equation}
is finite quantity which can be calculated with
Eq.~(\ref{eps_final}) numerically.

It follows from the time-reversal symmetry that clockwise and
counterclockwise polarized photons shift electron energy levels of
the ring with angular momenta $m$ and $-m$ equally. Indeed, the
eigenfrequencies (\ref{omega}) for clockwise and counterclockwise
circularly polarized photons are equal in the vacuum,
$\omega_{\mathbf{q},\eta,+}=\omega_{\mathbf{q},\eta,-}$. According
to Eq. (\ref{eps_final}), in this case we have the equality
$\varepsilon_{m,0}=\varepsilon_{-m,0}$ and the splitting (\ref{D})
vanishes. Therefore, the energy splitting needs the breaking of
the symmetry between virtual photons with different circular
polarizations. This can be achieved with filling the cavity with
an optically active medium, where the refractive indices $n_+$ and
$n_-$ are different. In what follows we will consider a metallic
quantum ring placed inside the cavity filled with such an
optically active medium. Let electron states with angular momenta
$m$ and $-m$ lie at the Fermi level $\mu$ of the ring when the
electron-photon interaction is absent (see Fig.~2a). Then,
summarizing in Eq.~(\ref{eps_final}) over states $m^\prime$ lying
over the Fermi level, we can obtain the vacuum-induced splitting
between otherwise degenerate states $m$ and $-m$ (see Fig.~2b). As
a result of the lifting of the degeneracy, the ground state of the
electron system in the ring possesses well-defined angular
momentum which corresponds to the nonzero electric current
\begin{equation}\label{jm}
j=\frac{m e \hbar}{2 \pi R^{2} m_{e}}.
\end{equation}

Since the current (\ref{jm}) is associated with the ground state,
it flows without any dissipation and is persistent. The
experimental observability of the vacuum-induced persistent
current depends on optimal choice of an optically active medium
filling the cavity, since the splitting (\ref{D}) depends on the
difference of the refractive indices, $\Delta n=|n_+-n_-|$ (see
Fig.~2c). For instance, the cavity can be filled with a
magnetogyrotropic medium based on ferrite garnets, where $\Delta n
\approx 5\times 10^{-3}$ (see Ref.~[\onlinecite{Ferrite}]). In
this case, the vacuum-induced splitting (\ref{D}) can be estimated
as $\Delta\varepsilon\sim 1~\mu$eV that is comparable to the value
of vacuum-induced Lamb shift in
atoms.\cite{Lamb_47,Bethe_47,Scully_10} The effect becomes even
more pronounced if the cavity is filled with an active media with
the circular dichroism \cite{CD_monography} or media based on a
metamaterial with a giant optical activity.\cite{Zheludev1} Then,
one of the two circularly polarized photon modes in the cavity is
suppressed and its contribution to the energy splitting (\ref{D})
can be neglected, which leads to the drastic increase of the
splitting. In this case, for $|m|\sim10^3$ the splitting is
$\Delta \varepsilon\sim 1$ meV (see Figs.~2d and 2e). Therefore,
the condition of observability of the vacuum-induced persistent
current, $\Delta \varepsilon\gg T$, can be easily satisfied at
liquid helium temperatures $T$.
\begin{figure}
\includegraphics[width=1.0\linewidth]{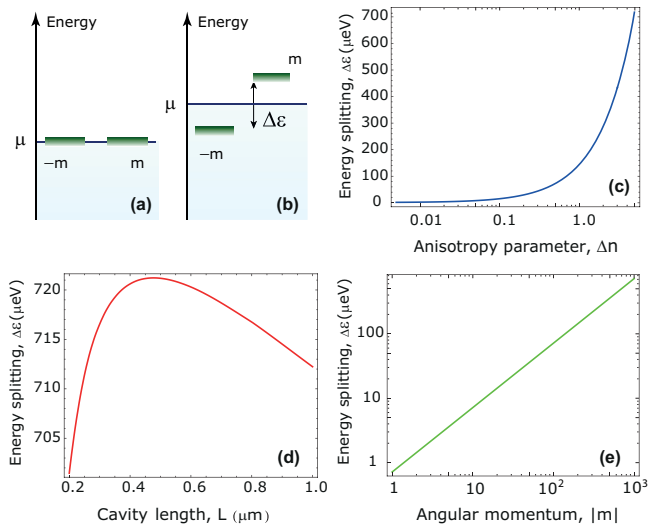}
\caption{(Color online) (a) Structure of energy levels close to
the Fermi level $\mu$ in the ring placed outside the cavity; (b)
Structure of energy levels close to the Fermi level $\mu$ in the
ring placed inside the cavity filled with an optically active
medium; (c) Vacuum-induced energy splitting for electron states
with $|m|=10^3$ as a function of the anisotropy parameter of the
optically active medium, $\Delta n = |n_+ - n_-|$, in the cavity
with $L=0.4~\mu$m; (d) Vacuum-induced energy splitting for
electron states with $|m|=10^3$ as a function of cavity length $L$
for the case of fully suppressed counterclockwise circularly
polarized mode; (e) Vacuum-induced energy splitting as a function
of angular momentum $|m|$ at the Fermi level of the ring for the
case of fully suppressed counterclockwise circularly polarized
mode in the cavity with $L=0.4~\mu$m.} \label{Fig2}
\end{figure}

To clarify the physical nature of the discussed effect, it should
be noted that the persistent current (\ref{jm}) arises from the
broken time-reversal symmetry in a chiral microcavity. Indeed, the
broken time-reversal symmetry leads to physical nonequivalence of
electron motion for mutually opposite directions in various
nanostructures: quantum wells,
\cite{Gorbatsevich_93,Aleshchenko_93,Omelyanovskii_96,Kibis_97,Kibis_98_1,Kibis_98_2,Kibis_99,Kibis_00,Diehl_07,Diehl_09,Kibis12}
quantum wires, \cite{Kibis11} carbon nanotubes,
\cite{Kibis_93,Kibis_01_1,Kibis_01_2} quantum rings,
\cite{Buttiker,Mailly,Kibis11} hybrid semiconductor-ferromagnet
nanostructures, \cite{Lawton_02} etc. As a result, a ground-state
current (persistent current) can exist in such nanostructures.
\cite{Buttiker,Mailly,Kibis11,Kibis12} Particularly, clockwise and
counterclockwise electron rotations in the quantum ring placed
inside the chiral microcavity are nonequivalent and, therefore,
the persistent current (\ref{jm}) appears.

For the ring with the radius $R\approx50$ nm and electron angular
momentum at the Fermi level $|m|\approx 1000$, the vacuum-induced
persistent current (\ref{jm}) can be estimated as $j\approx1$
$\mu$A. The magnetic field induced by the current can be detected
experimentally with a standard superconducting quantum
interference device (SQUID). In order to detect the current and to
exclude influence of the SQUID on the phenomenon, the SQUID should
be near a microcavity but outside it. Since the time-reversal
symmetry is broken in an optically active material filling the
microcavity, a built-in magnetic field can exist there. In order
to separate the magnetic field generated by the vacuum-induced
persistent current from other possible contributions,
difference-scheme measurements can be used. For instance,
magnetic-field measurements can be done for the microcavity with
two mirrors (where the vacuum-induced persistent current exists)
and for the same cavity with a removed mirror (where the
vacuum-induced persistent current is absent). Using of
compensation-scheme measurements
--- where the built-in magnetic field is compensated by an
opposite directed magnetic field --- is also possible.

The magnetic moment of a ring with the persistent current
(\ref{jm}) is given by
\begin{equation}\label{M}
M=\frac{m e \hbar}{2 m_{e}}.
\end{equation}
Due to the vacuum-induced magnetic moment (\ref{M}), the ring in
the cavity behaves as an artificial ``spin''. Replacing a single
ring with more complicated structure consisting of an array of
rings, which can be constructed experimentally,\cite{Zheludev2} we
will have an artificially designed Ising magnet. Thus, the
proposed structure forms a basis for the novel concept of optical
metamagnets which are expected to have intriguing properties. In
particular, it was recently demonstrated that resonator-based
systems with broken time-reversal symmetry can allow observation
of non-trivial topological phases of light.\cite{Yiddong} The
detail investigation of these effects, however, goes beyond the
scopes of the present paper and will be done elsewhere.\\

\section{Conclusion} Summarizing the aforesaid, we considered the
novel quantum electrodynamical effect emerging due to the
interaction of electrons in a quantum ring and electromagnetic
vacuum fluctuations in a resonator. We have shown that in the case
of the broken symmetry between clockwise and counterclockwise
circular polarizations of photon modes in the cavity, dressed
electronic states in the ring with opposite angular momenta are
split in energy. This vacuum-induced splitting leads to the
circulation of persistent current in the ring. Subsequently,
magnetic field generated by the persistent current can be detected
by SQUID techniques, that allows to claim the discussed phenomenon
as a first macroscopically observable vacuum effect in
nanostructures. As to possible applications of the effect to
devices, an array of quantum rings can be considered as a novel
type of metamaterial with magnetic properties (optical
metamagnet). It should be noted that the discussed effect is of
general character and will take place in any nanostructures which
are topologically homeomorphous to ring (particularly, in carbon
nanotubes).

\begin{acknowledgements}
The work was partially supported by FP7 IRSES (projects QOCaN and
SPINMET), Rannis ``Center of Excellence in Polaritonics'', and
RFBR project 13-02-90600. O.V.K. thanks the Nanyang Technological
University for the hospitality and O.K. acknowledges the support
from the Eimskip fund.
\end{acknowledgements}

\appendix
\section{Derivation of basic expressions}

In order to derive Eq.~(\ref{eps_final}) from Eq.~(\ref{eps}), we
need to find the matrix elements $\langle
m^{\prime},1_{\mathbf{q},\eta,1}|\hat{\cal{H}}_{int}|m,0\rangle$
and $\langle
m^{\prime},1_{\mathbf{q},\eta,2}|\hat{\cal{H}}_{int}|m,0\rangle$.
To reach this aim, we have to write the interaction Hamiltonian
(\ref{Hint}) in the elliptical polarization basis $\lambda=1,2$.
Using relations (\ref{Hopfield1})--(\ref{Hopfield2}) written in
the form
$\mathbf{e}_{1,\mathbf{q}}=\alpha_{\mathbf{q}}\mathbf{e}_{+}+
\beta_{\mathbf{q}}\mathbf{e}_{-}$ and
$\mathbf{e}_{2,\mathbf{q}}=\beta_{\mathbf{q}}\mathbf{e}_{+}-
\alpha_{\mathbf{q}}\mathbf{e}_{-}$, the electric field operators
of the cavity mode, $\hat{\mathbf{E}}_{\mathbf{q},\eta,\lambda}$,
can be written in the Hamiltonian (\ref{Hint}) as
\begin{widetext}
\begin{eqnarray}
\label{I}
\hat{\mathbf{E}}_{\mathbf{q},\eta,1}&=&i\sqrt{\frac{\hbar
\omega_{\mathbf{q},\eta,1}}{\epsilon_{0}L S}}\Big(
\hat{a}_{\mathbf{q},\eta,1} \alpha_{\mathbf{q}} e^{i
\mathbf{q}\cdot \mathbf{r}} \mathbf{e}_{+} +
\hat{a}_{\mathbf{q},\eta,1} \beta_{\mathbf{q}} e^{i
\mathbf{q}\cdot \mathbf{r}} \mathbf{e}_{-} -
\hat{a}^{\dagger}_{\mathbf{q},\eta,1} \alpha_{\mathbf{q}} e^{-i
\mathbf{q}\cdot \mathbf{r}} \mathbf{e}_{-} -
\hat{a}^{\dagger}_{\mathbf{q},\eta,1} \beta_{\mathbf{q}} e^{-i
\mathbf{q}\cdot \mathbf{r}} \mathbf{e}_{+} \Big) \sin \left(
\frac{\pi \eta z}{L} \right),\\
\label{II}
\hat{\mathbf{E}}_{\mathbf{q},\eta,2}&=&i\sqrt{\frac{\hbar
\omega_{\mathbf{q},\eta,2}}{\epsilon_{0}L S}}\Big(
\hat{a}_{\mathbf{q},\eta,2} \beta_{\mathbf{q}} e^{i
\mathbf{q}\cdot \mathbf{r}} \mathbf{e}_{+} -
\hat{a}_{\mathbf{q},\eta,2} \alpha_{\mathbf{q}} e^{i
\mathbf{q}\cdot \mathbf{r}} \mathbf{e}_{-}-
\hat{a}^{\dagger}_{\mathbf{q},\eta,2} \beta_{\mathbf{q}} e^{-i
\mathbf{q}\cdot \mathbf{r}} \mathbf{e}_{-} +
\hat{a}^{\dagger}_{\mathbf{q},\eta,2} \alpha_{\mathbf{q}} e^{-i
\mathbf{q}\cdot \mathbf{r}} \mathbf{e}_{+} \Big) \sin \left(
\frac{\pi \eta z}{L} \right),
\end{eqnarray}
\end{widetext}
where $\mathbf{e}_{\pm}=(\mathbf{e}_{x}\pm
i\mathbf{e}_{y})/\sqrt{2}$ are the unit vectors corresponding to
clockwise and counterclockwise circular polarizations of cavity
photons. Taking into account Eqs. (\ref{I})--(\ref{II}) and
keeping in mind that $\mathbf{e}_{+}\cdot \mathbf{t}(\varphi)=i
e^{i \varphi}/\sqrt{2}$ and $\mathbf{e}_{-}\cdot
\mathbf{t}(\varphi)=-i e^{-i \varphi}/\sqrt{2}$, the interaction
Hamiltonian (\ref{Hint}) reads as
\begin{widetext}
\begin{eqnarray}
\hat{\cal{H}}_{int}&=&-ieR\sum_{\mathbf{q},\eta} \left[
\sqrt{\frac{\hbar \omega_{\mathbf{q},\eta,1}}{2 \epsilon_{0}L S}}
\left( i \int \hat{a}_{\mathbf{q},\eta,1} \alpha_{\mathbf{q}} e^{i
\mathbf{q}\cdot \mathbf{r}} e^{i \varphi} d\varphi -i \int
\hat{a}_{\mathbf{q},\eta,1} \beta_{\mathbf{q}} e^{i
\mathbf{q}\cdot \mathbf{r}} e^{-i \varphi} d\varphi + i \int
\hat{a}^{\dagger}_{\mathbf{q},\eta,1} \alpha_{\mathbf{q}} e^{-i
\mathbf{q}\cdot \mathbf{r}} e^{-i \varphi} d\varphi\right.\right.\nonumber\\
&-&\left.i\int \hat{a}^{\dagger}_{\mathbf{q},\eta,1}
\beta_{\mathbf{q}} e^{-i \mathbf{q}\cdot \mathbf{r}} e^{i \varphi}
d\varphi \right) + \sqrt{\frac{\hbar \omega_{\mathbf{q},\eta,2}}{2
\epsilon_{0}L S}}\left( i \int \hat{a}_{\mathbf{q},\eta,2}
\beta_{\mathbf{q}} e^{i \mathbf{q}\cdot \mathbf{r}} e^{i \varphi}
d\varphi + i \int \hat{a}_{\mathbf{q},\eta,2} \alpha_{\mathbf{q}}
e^{i \mathbf{q}\cdot \mathbf{r}} e^{-i \varphi}
d\varphi\right.\nonumber\\
&+&\left.\left. i \int \hat{a}^{\dagger}_{\mathbf{q},\eta,2}
\beta_{\mathbf{q}} e^{-i \mathbf{q}\cdot \mathbf{r}} e^{-i
\varphi} d\varphi +i \int \hat{a}_{\mathbf{q},\eta,2}
\alpha_{\mathbf{q}} e^{-i \mathbf{q}\cdot \mathbf{r}} e^{i
\varphi} d\varphi \right)\right] \sin \left( \frac{\pi \eta z}{L}
\right). \label{III}
\end{eqnarray}
\end{widetext}
In what follows we will assume that the quantum ring is placed in
the center of the cavity ($z=L/2$). Consequently, the sine in the
last line of Eq.~(\ref{III}) can be omitted and the summation over
the index $\eta$ in Eq.~(\ref{III}) should be performed over odd
integer numbers. To proceed the derivation, we have to rewrite the
exponents $e^{\pm i \mathbf{q}\cdot \mathbf{r}}$ in
Eq.~(\ref{III}) using the polar coordinates
$\mathbf{r}=(R,\varphi)$ and $\mathbf{q}=(q,\theta)$. Then the
exponents can be written as $e^{\pm i \mathbf{q}\cdot \mathbf{r}}
= e^{\pm i q R \cos(\theta - \varphi)}$. Let us use the
Jacobi-Anger expansion \cite{Gradstein}
$$e^{i x \cos
\xi}=\sum_{n=-\infty}^{\infty}(i)^{n}J_{n}(x)e^{in\xi},$$ where
$J_{n}(x)$ is the Bessel function of the first kind. Then we
arrive to the expression $$e^{iqR\cos
(\varphi-\theta)}=\sum_{n=-\infty}^{\infty}(i)^{n}J_{n}(qR)e^{in(\varphi-\theta)}.$$
Using the well-known property of the Bessel function,
$J_{-n}(x)=(-1)^{n}J_{n}(x)$, the complex conjugation of this
exponent can be written as $$e^{-i q R \cos
(\varphi-\theta)}=\sum_{n=-\infty}^{\infty}(i)^{-n}J_{n}(qR)e^{in(\varphi-\theta)}.$$
As a result, the Hamiltonian (\ref{III}) takes the form
\begin{widetext}
\begin{eqnarray}
\hat{\cal{H}}_{int}&=&eR\sum_{\mathbf{q},\eta}\sum_{n=-\infty}^{\infty}
J_{n}(qR)e^{in\theta} \left[ \sqrt{\frac{\hbar
\omega_{\mathbf{q},\eta,1}}{2 \epsilon_{0}L S}} \left(
\hat{a}_{\mathbf{q},\eta,1} \alpha_{\mathbf{q}} (i)^{n}\int e^{-i
(n-1) \varphi} d\varphi - \hat{a}_{\mathbf{q},\eta,1}
\beta_{\mathbf{q}}
(i)^{n}\int e^{-i(n+1) \varphi} d\varphi\right.\right.\nonumber\\
&+&\hat{a}^{\dagger}_{\mathbf{q},\eta,1} \alpha_{\mathbf{q}}
(i)^{-n}\int e^{-i(n+1) \varphi} d\varphi-
\hat{a}^{\dagger}_{\mathbf{q},\eta,1} \beta_{\mathbf{q}}
(i)^{-n}\left.\int e^{-i(n-1) \varphi} d\varphi \right)+
\sqrt{\frac{\hbar \omega_{\mathbf{q},\eta,2}}{2 \epsilon_{0}L S}}
\left( \hat{a}_{\mathbf{q},\eta,2} \beta_{\mathbf{q}} (i)^{n}\int
e^{-i(n-1) \varphi} d\varphi\right.\nonumber\\
&+&\left. \hat{a}_{\mathbf{q},\eta,2} \alpha_{\mathbf{q}} (i)^{n}
\int e^{-i(n+1) \varphi} d\varphi
 + \hat{a}^{\dagger}_{\mathbf{q},\eta,2} \beta_{\mathbf{q}} (i)^{-n}\int e^{-i(n+1) \varphi}
 d\varphi + \hat{a}^{\dagger}_{\mathbf{q},\eta,2} \alpha_{\mathbf{q}}
(i)^{-n}\int e^{-i(n-1) \varphi} \left.d\varphi \right) \right].
\label{HintJA}
\end{eqnarray}
\end{widetext}
Performing in Eq.~(\ref{HintJA}) trivial integration over electron
angular coordinate $\varphi$, we arrive to the expression
\begin{widetext}
\begin{eqnarray}
\hat{\cal{H}}_{int}&=&eR\sum_{\mathbf{q},\eta}
\sum_{n=-\infty}^{\infty}J_{n}(qR)e^{in\theta}\left[
\sqrt{\frac{\hbar \omega_{\mathbf{q},\eta,1}}{2 \epsilon_{0}L S}}
\left( \hat{a}_{\mathbf{q},\eta,1} \alpha_{\mathbf{q}}
(i)^{n+1}\frac{e^{-i (n-1) \varphi}}{n-1} -
\hat{a}_{\mathbf{q},\eta,1} \beta_{\mathbf{q}} (i)^{n+1}
\frac{e^{-i (n+1) \varphi}}{n+1}\right.\right.\nonumber\\
&+&\left.\hat{a}^{\dagger}_{\mathbf{q},\eta,1} \alpha_{\mathbf{q}}
(i)^{-(n-1)}\frac{e^{-i (n+1) \varphi}}{n+1}-
\hat{a}^{\dagger}_{\mathbf{q},\eta,1} \beta_{\mathbf{q}}
(i)^{-(n-1)} \frac{e^{-i (n-1) \varphi}}{n-1}
\right)+\sqrt{\frac{\hbar \omega_{\mathbf{q},\eta,2}}{2
\epsilon_{0}L S}} \left( \hat{a}_{\mathbf{q},\eta,2}
\beta_{\mathbf{q}}
(i)^{n+1}\frac{e^{-i (n-1) \varphi}}{n-1}\right.\nonumber\\
&+& \left. \hat{a}_{\mathbf{q},\eta,2} \alpha_{\mathbf{q}}
(i)^{n+1} \frac{e^{-i (n+1) \varphi}}{n+1}+
\left.\hat{a}^{\dagger}_{\mathbf{q},\eta,2} \beta_{\mathbf{q}}
(i)^{-(n-1)}\frac{e^{-i (n+1) \varphi}}{n+1} +
\hat{a}^{\dagger}_{\mathbf{q},\eta,2} \alpha_{\mathbf{q}}
(i)^{-(n-1)}\frac{e^{-i (n-1) \varphi}}{n-1} \right) \right].
\label{HintJAint}
\end{eqnarray}
\end{widetext}
The matrix element of the Hamiltonian (\ref{HintJAint}) for
virtual photons with the polarization $\lambda=1$ is
\begin{widetext}
\begin{eqnarray}
\langle
m^{\prime},1_{\mathbf{q},\eta,1}|\hat{\cal{H}}_{int}|m,0\rangle
&=& eR \sqrt{\frac{\hbar \omega_{\mathbf{q},\eta,1}}{2
\varepsilon_{0}L S}}\sum_{n=-\infty}^{\infty}
(i)^{-(n-1)}J_{n}(qR)e^{in\theta}\left[
\frac{\alpha_{\mathbf{q}}}{n+1} \int\limits_{0}^{2 \pi}
\frac{d\varphi}{2
\pi} e^{i(m-m^{\prime}-n-1)\varphi} \right.\nonumber\\
& - &\left.\frac{\beta_{\mathbf{q}}}{n-1} \int\limits_{0}^{2 \pi}
\frac{d\varphi}{2 \pi} e^{i(m-m^{\prime}-n+1)\varphi} \right].
\label{U1}
\end{eqnarray}
\end{widetext}
The integration over the angular coordinate $\varphi$ in
Eq.~(\ref{U1}) gives the Kronecker deltas
$\delta_{n,m-m^{\prime}-1}$ and $\delta_{n,m-m^{\prime}+1}$, which
reduce the summation over the index $n$ in Eq.~(\ref{U1}) to the
single term:
\begin{widetext}
\begin{eqnarray}
\langle
m^{\prime},1_{\mathbf{q},\eta,1}|\hat{\cal{H}}_{int}|m,0\rangle
=-eR \sqrt{\frac{\hbar \omega_{\mathbf{q},\eta,1}}{2
\varepsilon_{0}L
S}}\frac{(i)^{-(m-m^{\prime})}}{m-m^{\prime}}e^{i(m-m^{\prime})\theta}
\left[ \alpha_{\mathbf{q}} J_{m-m^{\prime}-1}(qR)
e^{-i\theta}+\beta_{\mathbf{q}} J_{m-m^{\prime}+1}(qR)e^{i\theta}
\right]. \label{U1_2}
\end{eqnarray}
\end{widetext}
Deriving the matrix element of the interaction Hamiltonian
(\ref{HintJAint}) for virtual photons with the polarization
$\lambda=2$ in the same way, we arrive to the expression
\begin{widetext}
\begin{eqnarray}
\langle
m^{\prime},1_{\mathbf{q},\eta,2}|\hat{\cal{H}}_{int}|m,0\rangle =
-eR \sqrt{\frac{\hbar \omega_{\mathbf{q},\eta,2}}{2
\varepsilon_{0}L
S}}\frac{(i)^{-(m-m^{\prime})}}{m-m^{\prime}}e^{i(m-m^{\prime})\theta}
\left[ \beta_{\mathbf{q}}J_{m-m^{\prime}-1}(qR)e^{-i\theta}
-\alpha_{\mathbf{q}} J_{m-m^{\prime}+1}(qR) e^{i\theta} \right].
\label{U2_2}
\end{eqnarray}
\end{widetext}
Substituting Eqs.~(\ref{U1_2})--(\ref{U2_2}) into Eq.~(\ref{eps})
and passing from summation over photon wave vectors $\mathbf{q}$
to integration, $\sum_{\mathbf{q}} \rightarrow S/(2 \pi)^{2}
\int_{0}^{\infty} q dq \int_{0}^{2\pi} d\theta$, we arrive to
Eqs.~(\ref{eps_final})--(\ref{D}) which are the basic expressions
for the analysis of the discussed effect.

\end{document}